\documentclass{ws-procs9x6}
\usepackage{feyn}
\usepackage{MnSymbol}

\newcommand{\refeq}[1]{(\ref{#1})}

\newcommand{\feynslash}[1]{#1\kern-0.45em/}

\begin{document}

\title{FINITE ONE-LOOP RADIATIVE CORRECTIONS IN THE
LORENTZ- AND CPT-VIOLATING QED EXTENSION}

\author{D.H.T.\ FRANCO and A.H.\ GOMES$^*$}

\address{Universidade Federal de Vi\c{c}osa,
	 Vi\c{c}osa, 36570-000, Minas Gerais, Brazil\\
$^*$E-mail: andre.herkenhoff@ufv.br}

\begin{abstract}
We report on our current progress in the study of finite
one-loop radiative corrections in the minimal Lorentz-
and CPT-violating QED extension.
\end{abstract}

\bodymatter

\section{Introduction}

Quantum gravity effects coming from Planck scale energies
may appear as small violations of fundamental laws in the
limit of low energies. One possibility is the breaking of
both Lorentz and CPT symmetries. A comprehensive investigation
of this case is made viable, for instance, in the framework
known as the Standard Model Extension\cite{sme}, the most general
effective model comprising particle Lorentz and CPT
violation but coordinate invariant. The minimal extension
respects the Standard Model $SU(3) \otimes SU(2) \otimes U(1)$
gauge symmetry and is power-counting renormalizable. In this
proceedings contribution we report our progress in the study of finite
one-loop radiative corrections in the minimal Lorentz- and
CPT-violating QED extension.

\section{The QED Extension and setup for one-loop evaluations}
 
The Lorentz- and CPT-violating QED extension for one Dirac fermion
can be described by the Lagrange density\cite{sme}
\begin{eqnarray}\label{qedex}
\mathcal{L} & = & i\, \overline{\psi}(\gamma^{\,\mu} +
\Gamma_1^\mu)D_\mu\,\psi - \overline{\psi}(m+M_1)\psi -
\frac{1}{4}F^{\mu\nu} F_{\mu\nu}	\nonumber \\
& & - \frac{1}{4}(k_F)_{\mu\nu\rho\sigma} F^{\mu\nu}F^{\rho\sigma}
+ \frac{1}{2}(k_{AF})^\mu \epsilon_{\mu\nu\rho\sigma} A^\nu F^{\rho\sigma},
\end{eqnarray}
where Lorentz violation (LV) is parameterized by coefficients
in the form of constant background fields --- see the last two
terms of Eq.\ \refeq{qedex} and the definitions
\begin{equation}
\Gamma_1^\mu \equiv c^{\,\lambda\mu}\gamma_\lambda + 
d^{\lambda\mu}\gamma_5\gamma_\lambda + e^{\,\mu} + 
if^\mu\gamma_5 + \frac{1}{2}\,g^{\kappa\lambda\mu}\sigma_{\kappa\lambda},
\end{equation}
\begin{equation}
M_1 \equiv a^{\,\mu}\gamma_\mu + b^{\,\mu}\gamma_5\gamma_\mu + 
\frac{1}{2}\,H^{\mu\nu}\sigma_{\mu\nu}.
\end{equation}
We also add to \refeq{qedex} a photon mass term $\frac{1}{2}\,\lambda^2 A^2$
to control infrared divergences and a gauge-fixing piece
$-\frac{1}{2\alpha}(\partial_\mu A^\mu)^2$.

We follow the guidelines presented in Ref.\ \refcite{one-loop-renorm} 
to deal with one-loop radiative corrections: evaluations are performed in the 
so-called concordant frames to avoid issues with the perturbative expansion; 
nonlinear LV is disregarded because it could lead to effects comparable to 
multiloop ones, which we do not consider; 
and LV pieces of the Lagrange density \refeq{qedex} are treated 
as new interaction vertices, effectively entering as propagator 
or vertex insertions, as allowed by the experimental smallness of LV.

\section{Results and discussion}

\begin{figure}
\begin{center}
\psfig{file=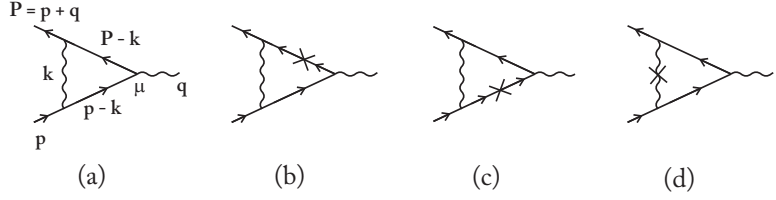,width=4.0in}
\end{center}
\caption{(a) Conventional QED vertex correction and 
		(b)-(d) Lorentz-violating insertions.}
\label{fig:fig1}
\end{figure}

As a first step, we considered only the $b_\mu$ and $(k_{AF})_\mu$ 
coefficients. Results for the photon self-energy can be found in 
Ref.\ \refcite{sme}, so our goal was first the vertex correction 
diagram, Fig.\ \ref{fig:fig1}(a). The electron self-energy 
was used so far only to check the related gauge identity (more details below), 
and its evaluation will be our next goal. 
The LV contribution with one $b_\mu$ 
insertion comes from two diagrams, Figs.\ \ref{fig:fig1}(b) and 
\ref{fig:fig1}(c), with the new Feynman rule 
\begin{equation}
\feyn{f a f x f a f} = - i b_\mu \gamma_5 \gamma^\mu,
\end{equation}
and the contribution from $(k_{AF})_\mu$ comes from the one in 
Fig.\ \ref{fig:fig1}(d), where
\begin{equation}
\mu\,\,\, \feyn{ g g } \hspace{-.8cm}\boldsymbol{\bigtimes} 
\quad\,\,\,\,\,\nu = 2 (k_{AF})^\alpha \varepsilon_{\alpha\mu\beta\nu} 
k^\beta.
\end{equation}
These insertions lower the superficial degree of divergence of 
vertex-correction loop integrals, making them actually finite, avoiding 
complications related to $\gamma_5$.

Aiming for generality, we computed both cases with fermions and 
photon off shell, i.e., 
$P^2\neq m^2\neq p^2$ and $q^2\neq 0$, and we made no use of the 
Dirac equation. Since Gordon identities are unavailable to 
make further simplifications --- they are valid only on shell 
--- the final result itself is not very illuminating because of 
its size and structure, but it is worth mentioning that because 
we do not use $P^2=m^2=p^2$ our expressions contain not only 
pieces symmetric under $P\leftrightarrow{p}$ but also pieces 
that are antisymmetric. On the other hand, symmetry under $P\leftrightarrow{p}$ 
is recovered when the fermions go on shell, and the final expressions are 
greatly simplified. For instance, the LV correction due to 
$(k_{AF})_\mu$ has the structural form
\begin{align}
\Gamma^\mu_{k_{AF}} \sim \Big\{ & \feynslash{k}_{AF}\gamma_5(P+p)^\mu,\,\,\,
 \gamma_5\sigma^{\mu\nu}(P+p)_\nu k_{AF}\cdot(P+p),	\nonumber	\\
&	 \gamma^\mu\gamma_5 k_{AF}\cdot(P+p),\,\,\, 
\gamma_5\sigma^{\kappa\lambda}(k_{AF})_\kappa(P+p)_\lambda(P+p)^\mu \Big\},
\end{align}
and the result from the $b^\mu$ insertion has the same structure 
but with the additional form $\gamma_5\sigma^{\mu\nu}b_\nu$, which 
vanishes for the $(k_{AF})_\mu$ contribution.

As a consistency check, we verified the Ward-Takashi identity, 
$q_\mu\Gamma^\mu=\Sigma(p+q)-\Sigma(p)$, for both cases up to order 
$q^2$ with arbitrary IR regulator $\lambda$, but so far with fermions 
on shell --- the off shell case is still to be done. The Ward-Takashi identity
demands the evaluation of LV contributions to the electron self energy, which 
are divergent. Nevertheless, only the finite piece of this integral 
is momentum dependent and contributes to the right-hand side of the 
Ward-Takashi identity. 
It is interesting to notice that, different from the conventional QED 
case, $q_\mu \Gamma^\mu \neq 0$ even when the Dirac equation is used for 
the external leg spinors: even in this situation the momentum shift in 
the electron self energy still contributes to the identity.

The next step is to establish 
the finite contributions coming also from the divergent 
piece of the electron self energy, which may require a careful choice of 
subtraction points. Applications in physical processes will also be 
envisaged, as well as the analysis for other LV coefficients.

\section*{Acknowledgments}

Financial support was provided by CNPq (Brazil).




\begin{thebibliography}{x}
\bibitem{sme}
D.\ Colladay, and V.A.\ Kosteleck\'{y},
Phys.\ Rev.\ D {\bf 58}, 116002 (1998).
\bibitem{one-loop-renorm}
V.A.\ Kosteleck\'{y}, C.D.\ Lane, and A.G.M.\ Pickering, 
Phys.\ Rev.\ D {\bf 65}, 056006 (2002).			
\end{thebibliography}
\end{document}